\documentclass[american,aps,pra,superscriptaddress,twocolumn]{revtex4-2}

\usepackage[T1]{fontenc}
\usepackage[sc]{mathpazo}
\usepackage{amsmath}
\usepackage{amssymb}
\usepackage{enumerate}
\usepackage{amsthm}% theorems and proofs
\usepackage{color}
\usepackage{graphicx} % For including graphics N.B. pdftex graphics driver

\usepackage{hyperref}
\hypersetup{pdfpagemode=UseNone}

\newcommand{\ket}[1]{|#1\rangle}

\newcommand{\braket}[1]{\langle#1\rangle}

\begin{document}

\title{Triple measurements uncertainty and the distinguishment between the separable and entangled states}

 \author{Minyi Huang}
 \email{11335001@zju.edu.cn}
 \email{hmyzd2011@126.com}
 %\email{hmyzstu2011@126.com}
 \affiliation{Department of Mathematical Sciences, Zhejiang Sci Tech University, Hangzhou 310007, PR~China}
% \affiliation{Interdisciplinary Center of Quantum Information, State Key Laboratory of Modern Optical Instrumentation, and Zhejiang Province Key Laboratory of Quantum Technology and Device, Department of Physics, Zhejiang University, Hangzhou 310027, China}

\author{Ray-Kuang Lee}
 \email{rklee@ee.nthu.edu.tw}
\affiliation{Department of Physics, National Tsing Hua University, Hsinchu 300, Taiwan}
\affiliation{Institute of Photonics Technologies, National Tsing Hua University, Hsinchu 300, Taiwan}

\begin{abstract}
Uncertainty and entanglement are both profound and key concepts in quantum theory.
For three observables, the tightest uncertainty constants for both product and summation forms are revealed.
In this work, we give an alternative proof for three observables, also with a physical interpretation
of the uncertainty constants. Our results show that such constants are intimately connected with the
distinguishment between separable and entangled states.
\end{abstract}

\maketitle

\section{Introduction}
\label{sec:intro}
%=============================================================================%
Uncertainty is one of the most fundamental concepts in quantum mechanics. In 1927,
Heisenberg discovered his seminal uncertainty principle which defines the fundamental constraints on quantum
measurements \cite{heisenberg1927anschaulichen}. In 1929, Robertson formulated precise uncertainty
relations as symmetric quadratic polynomials of the variances of two observables \cite{robertson1929uncertainty}.
For any two quantum mechanical observables $A$ and $B$, Robertson has found the following  uncertainty inequality,
\begin{equation}
\Delta A\Delta B\geqslant \frac{1}{2}|\braket{A,B}|, \label{uncertain}
\end{equation}
where $\Delta \Omega\triangleq \sqrt{\braket{\Omega^2}-\braket{\Omega}^2}$ is the standard deviation and $\braket{\Omega}$ is the expected value of the observable $\Omega$ with respect to the state $\rho$.
The variance is denoted as $\Delta \Omega^2\triangleq \braket{\Omega^2}-\braket{\Omega}^2$. Since then, there are lots of discussions in this field, generalizing the uncertainty relation to more observables and different metrics \cite{robertson1934indeterminacy,maassen1988generalized,arthurs1988quantum,ozawa2004uncertainty,busch2007heisenberg,
wehner2008higher,wehner2010entropic,watanabe2011uncertainty,huang2012variance,chen2016variance,li2016equivalence,coles2017entropic,
miyadera2008heisenberg,dodonov2019uncertainty}.

 In the case of three observables,  an investigation of the uncertainty relation reveals the tightest uncertainty constants for both product and summation forms~\cite{liang2024signifying}. For three observables $H_j$, it is showed that
 \begin{eqnarray}
&&\Pi_{j=1}^{3}\Delta H_j^2\geqslant (\frac{1}{\sqrt{3}})^3(\Pi_{j=1}^{3}|\braket{[H_j,H_{j+1}]}|,\label{varpro}\\
&&\sum_{j=1}^{3}\Delta H_j^2\geqslant\frac{1}{\sqrt{3}}\sum_{j=1}^{3}|\braket{[H_j,H_{j+1}]}|. \label{var2}
\end{eqnarray}

In addition to the variance, entanglement also gives the most important concepts in quantum information theory. The discussions of entanglement can be dated back to the Einstein-Podolsky-Rosen paradox~\cite{einstein1935can}. Since then, entanglement has been one of the key issues in quantum
theory, which has greatly deepen our understanding  of quantum mechanics, with numerous applications in quantum technologies \cite{horodecki2009quantum}.
Entanglement also stimulates the discussions on the constraints of probability distributions, operator structures, game theory and many other problems in the field of mathematical physics~ \cite{bell1964einstein,cirel1980quantum}.

Behind uncertainty and entanglement lies the concept of non-commutativity of operators. In fact, one can see immediately that the uncertainty in Eq.~(\ref{uncertain}) comes from the non-commutativity of observables. On the other hand, researchers have showed the crucial role of non-commutativity in the discussions of entanglement or nonlocality~\cite{bell1964einstein,cirel1980quantum,landau1987experimental,landau1987violation}.
 Thus it is natural that researches have shown close relations between the two concepts of uncertainty and entanglement~\cite{hofmann2003violation,wolf2009measurements,oppenheim2010uncertainty}.
In this paper, we follow a different route to prove the inequalities of the variance in the sum and product form. Our proof gives a more natural physical meaning, with a clear physical interpretation to the uncertainty constants. It is showed that they are intimately related to the distinguishment between the separable and entangled states.

The structure of the paper is organized as follows. Section II is the proof of the inequalities of the variance, i.e. Eqs.~(\ref{varpro}) and (\ref{var2}).
In Section III, it is shown that these inequalities, as well as the uncertainty constants,
can be utilized to give a way to distinguish between separable and entangled states. In Section IV, a concrete example is given. Section V is some discussions and the summary. Section VI is the appendix.

\section{The inequalities of variance in the sum and product forms}

In order to prove the uncertainty relations in Eqs.~(\ref{varpro}) and (\ref{var2}), one can introduce an ancillary system and consider the variance of the global system to indicate the local uncertainty relation in the subsystem. Then, the sum form in Eq.~(\ref{var2}) suggests that it may be more suitable for giving a physical interpretation. For this reason, unlike Ref.~\cite{liang2024signifying}, we first prove Eq.~(\ref{var2}) instead of Eq. (\ref{varpro}). As a consequence,  Eq.~(\ref{varpro}) is directly derived from  Eq.~(\ref{var2}).
%Let $a, b, c$ and $a', b', c'$ be observables. be three

Now, we define a global operator $R=\sum_{j=1}^3 H_j \otimes \sigma_j$, where $\sigma_j$ are the Pauli operators.
Direct calculations show that

%\begin{eqnarray}
%\nonumber R^2&=&(H_1^2+H_2^2+H_3^2)\otimes I+[H_1,H_2]\otimes i\sigma_3\\
%&+&[H_2,H_3]\otimes i\sigma_1+[H_3,H_1]\otimes i\sigma_2.
%\end{eqnarray}

\begin{equation}
R^2=\sum_{j=1}^{3} H_j^2\otimes I+\sum_{j=1}^{3}[H_j,H_{j+1}]\otimes i\sigma_{j+2}
\end{equation}
Note that the subscript $j$  modulo by $3$, i.e., $j$ mod $3$.

Now we want to investigate the variance of $R$. To this end, we first consider the bound of $R$ and its square.
Generally, it depends on the concrete form of $R$ and $R^2$.
Suppose that the state is a product state, i.e. $\mu\otimes\nu$, where $\mu$ and $\nu$ are states of the subsystems.

Then we have
\begin{eqnarray}
&&Tr (\mu\otimes\nu)R^2=\sum_{j=1}^{3}\braket{H_j^2}+\sum_{j=1}^{3}\braket{i[H_j,H_{j+1}]}\braket{ \sigma_{j+2}},\label{square2}\\
&&[Tr (\mu\otimes\nu)R]^2=(\sum_{j=1}^{3} \braket{H_j}\braket{\sigma_j})^2.
\end{eqnarray}

Note that by the Schwarz inequality, for any state $\rho$ , one has
\begin{equation}
[Tr \rho R]^2 \leqslant Tr \rho R^2.\label{varrho}
\end{equation}

For a product state, still by the Schwarz inequality, we have
\begin{eqnarray*}
[Tr (\mu\otimes\nu)R]^2&=&(\sum_{j=1}^{3} \braket{H_j}\braket{\sigma_j})^2\\
&\leqslant& (\sum_{j=1}^{3} \braket{H_j}^2)(\sum_{j=1}^{3}\braket{\sigma_j}^2).
\end{eqnarray*}
Note that $\sum_j \braket{\sigma_j}^2\leqslant 1$ \cite{hofmann2003violation,ma2017experimental}, thus
\begin{equation}
[Tr (\mu\otimes\nu)R]^2\leqslant \sum_{j=1}^{3} \braket{H_j}^2.\label{Hje2}
\end{equation}

By Eq. (\ref{square2}) and Eq. (\ref{Hje2}), one can see that
\begin{eqnarray}
\nonumber \Delta [(\mu\otimes\nu)R]^2&\triangleq&Tr (\mu\otimes\nu)R^2-[Tr (\mu\otimes\nu)R]^2\\
&\geqslant&\sum_{j=1}^{3} \Delta H_j^2 +\sum_{j=1}^{3}\braket{i[H_j,H_{j+1}]}\braket{ \sigma_{j+2}}\label{var}
\end{eqnarray}
Note that identity holds in Eq. (\ref{var}) only if identity holds in Eq. (\ref{Hje2}).
According to the Schwarz inequality, it is valid
when $\braket{H_j}=t\braket{\sigma_j}$ and $|\braket{\sigma_j}|=\frac{1}{\sqrt{3}}$.
Moreover, according to the appendix in Section VI, actually the state $\nu$ can be specifically chosen to satisfy
\begin{equation}
\sum_{j=1}^{3}\braket{i[H_j,H_{j+1}]}\braket{ \sigma_{j+2}}=-\frac{1}{\sqrt{3}}\sum_{j=1}^{3}|\braket{[H_j,H_{j+1}]}|. \label{bound}
\end{equation}

Thus when $\braket{H_j}=t\braket{\sigma_j}$ and $|\braket{\sigma_j}|=\frac{1}{\sqrt{3}}$,
 it follows from Eqs. (\ref{varrho}) and (\ref{bound}) that,
\begin{equation}
\sum_{j=1}^{3}\Delta H_j^2\geqslant\frac{1}{\sqrt{3}}\sum_{j=1}^{3}|\braket{[H_j,H_{j+1}]}|\label{varsum}.
\end{equation}
In particular, one can see immediately that
 Eq. (\ref{varsum}) is valid when $\braket{H_j}=0$. Now for general $H_j$, one can take $H_j'=H_j-\braket{H_j}I$, then $\braket{H_j'}=0$. However, since
$\Delta H_j'=\Delta H_j$ and $[H_j',H_{j+1}']=[H_j,H_{j+1}]$, Eq. (\ref{varsum}) is valid for any matrices $H_j$. Thus we complete the proof of Eq. (\ref{var2}).

By Eq. (\ref{varsum}), one can further obtain the triple measurements uncertainty in the product form.
In fact, by the Algebraic-Geometric Mean inequality,
\begin{eqnarray}
\nonumber\frac{1}{3}\sum_{j=1}^{3}\Delta H_j^2&\geqslant&\frac{1}{\sqrt{3}}(\frac{1}{3}\sum_{j=1}^{3}|\braket{[H_j,H_{j+1}]}|)\\
&\geqslant& \frac{1}{\sqrt{3}}(\Pi_{j=1}^{3}|\braket{[H_j,H_{j+1}]}|)^{\frac{1}{3}}.
\end{eqnarray}
It follows that
\begin{eqnarray}
(\frac{1}{3}\sum_{j=1}^{3}\Delta H_j^2)^3\geqslant (\frac{1}{\sqrt{3}})^3\Pi_{j=1}^{3}|\braket{[H_j,H_{j+1}]}|.
\end{eqnarray}
Apparently, when $\Delta H_j^2$ are equal, one can deduce from the above equation that
\begin{eqnarray}
\Pi_{j=1}^{3}\Delta H_j^2\geqslant (\frac{1}{\sqrt{3}})^3\Pi_{j=1}^{3}|\braket{[H_j,H_{j+1}]}|.\label{varpro2}
\end{eqnarray}
Now for the case $\Delta H_j^2$ are not equal, one can choose coefficients
$\kappa_j= \frac{(\Pi_{k=1}^{3}\Delta H_k^2)^\frac{1}{6}}{\Delta H_j}$ and $H_j'=\kappa_j H_j$. Now direct calculations show
that $\Pi_{j=1}^{3} \kappa_j^2=1$ and so that $(\Delta H_j')^2$ are equal.
 Moreover, $\Pi_{j=1}^{3}(\Delta H_j')^2=\Pi_{j=1}^{3}\Delta H_j^2$ and
$\Pi_{j=1}^{3}|\braket{[H_j',H_{j+1}']}|=\Pi_{j=1}^{3}|\braket{[H_j,H_{j+1}]}|$. Thus one can see that
Eq. (\ref{varpro2}) is also valid for general $H_j$, which completes the proof of Eq. (\ref{varpro}).\\

\section{The difference between separable and entangled states}

In this section, we show that the above discussions can be related to the distinguishment between separable and entangled states. Recall that a state is said to be separable if it can be written as a convex combination of the product states, i.e.
\begin{equation}
\rho=\sum_j \lambda_j \mu_j\otimes \nu_j.\label{sep}
\end{equation}
By the Schwarz inequality, one can verify that
\begin{equation}
\Delta \rho^2\geqslant\sum_j \lambda_j \Delta(\mu_j\otimes \nu_j)^2.
\end{equation}
Note that if the variance of any product state is larger than zero, i.e., $\Delta(\mu_j\otimes \nu_j)^2> 0$, then for any
separable state $\rho$, $\Delta\rho^2>0$. Since the separable states form a compact set \cite{watrous2018theory}, one can
obtain that there exists some positive constant $c>0$ such that $\Delta\rho^2\geqslant c>0$ for any separable state $\rho$.
 Now note that in the derivation above, we have used product state. If the identity in Eqs. (\ref{var}) and (\ref{varsum})
 does not hold for any product state,
then the variance of any product state is larger than zero. Thus we know that $\Delta\rho^2\geqslant c>0$ for
 any separable state $\rho$. On the other hand, the variance of a general state can violate such a bound. In fact, if $\rho$ is
 an eigenstate of $R=\sum_{j=1}^3 H_j\otimes \sigma_j$, then its variance $\Delta\rho^2=0$. Moreover, since it is not separable,
 this eigenstate must be an entangled state.
  Thus one can obtain a sufficient condition of entanglement, if the variance of the state is less than the constant $c$,
  it is an entangled state. \\

\section{An example}
In this section, we give an example to show the largest bound of $R^2$ and how it can be used in the distinguishment between
separable and entangled states. For convenience, we assume that $H_i^2=I$.
In this case, $R^2$ reduces to
\begin{equation}
R^2=3 I+\sum_{j=1}^{3}[H_j,H_{j+1}]\otimes i\sigma_{j+2}.
\end{equation}
Now the norm of $R^2$ is
\begin{eqnarray}
\nonumber \|R^2\|&=&\|3I+\sum_{j=1}^{3}[H_j,H_{j+1}]\otimes i\sigma_{j+2}\|\\
\nonumber &\leqslant& 3+\sum_{j=1}^{3}\|[H_j,H_{j+1}]\otimes i\sigma_{j+2}\| \\
&\leqslant& 9. \label{norm}
\end{eqnarray}
Note that Eq. (\ref{norm}) is due to the fact that $\|H_j\|=1$ and thus $\|[H_j,H_{j+1}]\|\leqslant 2$.
It follows that
\begin{eqnarray}
&&Tr \rho R^2 \leqslant9, \label{bound2}\\
&&|Tr \rho R |\leqslant3. \label{bound3}
\end{eqnarray}
In deed, there exist some $H_j$ and $\rho$ such that the upper bounds in Eqs. (\ref{bound2}) and (\ref{bound3}) can be saturated.
Note that $\|[H_j,H_{j+1}]\otimes i\sigma_{j+2}\|$ is the maximal absolute eigenvalue $\lambda(max)$ of
$[H_j,H_{j+1}]\otimes i\sigma_{j+2}$.
Moreover, for any state
 $\phi$, $\braket{\phi|\sum_{j=1}^{3}[H_j,H_{j+1}]\otimes i\sigma_{j+2}|\phi}\leqslant \sum_{j=1}^{3}\|[H_j,H_{j+1}]\otimes
 i\sigma_{j+2}\|=\sum_{j=1}^{3}\lambda(max)_j$. Thus
to obtain the maximal value in Eq. (\ref{norm}) or Eq. (\ref{bound2}), the operators $[H_j,H_{j+1}]\otimes i\sigma_{j+2}$
necessarily have a common eigenstate. To this end, one can assume that the operators $[H_j,H_{j+1}]\otimes i\sigma_{j+2}$ commute. Direct calculations show that
this is valid when $[H_j, H_{j+1}]$ anti-commute. Thus they can generate a Clifford algebra. For example, one can take
$H_j=\sigma_j$, then
\begin{equation}
R=\sum_{j=1}^3 \sigma_j\otimes \sigma_j.\label{example}
\end{equation}
The maximal eigenvalue of $R^2$ is $9$, with the corresponding eigenvector $\frac{1}{\sqrt{2}}(0,-1,1,0)^T$ which saturates the upper bound. Note that $\frac{1}{\sqrt{2}}(0,-1,1,0)^T$ is a pure state and can be written as
$\frac{1}{\sqrt{2}}(-\ket{01}+\ket{10})$, whose Schmidt number is two. Thus it is an entangled state.

On the other hand, when $\rho=\mu\otimes\nu$ be a product state, since $\sum_j \braket{\sigma_j}^2\leqslant 1$, then by the Schwarz inequality we have
\begin{eqnarray}
\nonumber Tr (\mu\otimes\nu)R^2&=&3+\sum_{j=1}^{3}\braket{i[H_j,H_{j+1}]}\braket{ \sigma_{j+2}}\\
\nonumber &\leqslant& 3+\sqrt{\sum_{j=1}^{3}\|[H_j,H_{j+1}]\|^2}\\
&\leqslant& 3+2\sqrt{3}.
\end{eqnarray}
Hence
\begin{equation}
|Tr (\mu\otimes\nu)R |\leqslant \sqrt{3+2\sqrt{3}}.\label{bound4}
\end{equation}
The above bound is also valid for separable states.

In \cite{hofmann2003violation}, the researchers propose a method to detect entanglement by considering the variance of
separable and entangled states, when the observable can be written as $\sum_j A_j\otimes I+I\otimes B_j$.
Now since we have obtained the values above, such a distinguishment between separable and entangled state can also be made in our case by comparing the variance. In fact, by taking
$H_j=\sigma_j$ and one cannot obtain an identity in Eq. (\ref{var2}). Thus the variance in Eq. (\ref{var}) is strictly larger
than zero. Thus for the separable state, the variances have some lower bound $c>0$ which can be violated by the entangled states. That is, if the variance is less than such constant $c$, the state is entangled. However, it is usually difficult to calculate
such a constant. Instead, one can consider the expected values in Eqs. (\ref{bound3}) and (\ref{bound4}). According to the discussions above, one can see that if the expected value is larger than $\sqrt{3+2\sqrt{3}}$, then the state is entangled.

\section{Discussions}
The proof of the uncertainty relations in our paper is inspired by the work \cite{liang2024signifying}. However, the proof therein is based on the positivity of the expected values of the observables. First proving the product form of uncertainty, the sum form of uncertainty can be obtained similarly with a previous used lemma \cite{liang2024signifying}.
We follow a different route to directly consider the variance and first prove the sum form of uncertainty relation. Such a way has a clear physical meaning. Correspondingly, the way of mathematical deduction is different from \cite{liang2024signifying}.
This difference is also partly reflected by the establishment of the product form of uncertainty as a corollary of the sum form.

In Ref.~\cite{hofmann2003violation}, the violation of local uncertainty relations can be viewed as a signature of entanglement. The discussions therein mainly focused on the case where the
global observable can be written as $\sum_j A_j\otimes I+I\otimes B_j$.
The key is the commutativity of $A_j\otimes I$ and $I\otimes B_j$. In such a case, the variance
is additive, i.e. $\Delta (\sum_j A_j\otimes I+I\otimes B_j)^2=\sum_j (\Delta A_j^2+\Delta B_j^2)$. This good
property gives a natural lower bound of the variance separable states. In the general case, the complicated form
of the observable, e.g. $\sum_j A_j\otimes C_j+D_j\otimes B_j$
 usually does not give such a neat formula of uncertainty directly, due to the fact that $A_j\otimes C_j$ and $D_j\otimes B_j$
 does not commute in general. However, for the observable $R=\sum_{j=1}^3 H_j \otimes \sigma_j$ considered in this paper,
with the fine properties of Pauli matrices, it is still possible to obtain
 a lower bound of the variance without the commutativity of
$H_j \otimes \sigma_j$.

Moreover, it is demonstrated that in section IV that the variance of states can help in the
distinguishment between separable and entangled states. Such a discussion naturally generalize the results
in Ref.~\cite{hofmann2003violation} to the case of a observable without enough commutative property.
Furthermore, we showed that the coefficient of uncertainty obtained in
\cite{liang2024signifying} is essentially related to the distinguishment between separable and entangled states. This gives a
physical interpretation to the result in \cite{liang2024signifying}. On the other hand, as was pointed out in \cite{hofmann2003violation},
it is usually difficult to determine the value of the lower bound of the variance. For this reason, we also showed that
the expected value of $R=\sum_{j=1}^3 H_j \otimes \sigma_j$ can be used in the distinguishment task, which is usually easier in
manipulation and calculation.

Our discussions may shed light on the discussions of the relations of entanglement, uncertainty and the
algebraic properties of the observables without enough commutativity.

\section{Appendix}
By choosing some appropriate $\nu$, one can ensure that the identities hold in Eqs. (\ref{Hje2}), (\ref{var}) and (\ref{bound}) simultaneously when $\braket{H_j}=t\braket{\sigma}_j$.
Note that since $i[H_j,H_{j+1}]$ are Hermitian, to ensure that $\braket{i[H_j,H_{j+1}]}\braket{\sigma_{j+1}}$ is negative, we only need to
select some state $\ket{\phi}$ such that $\braket{\sigma_{j+1}}$ has the opposite sign to  $\braket{i[H_j,H_{j+1}]}$. There are eight possible cases, for which the state are listed as follows. Calculations verify the signs and that $|\braket{\sigma_{j}}|=\frac{1}{\sqrt{3}}$.

1. $\{+,+,+\}$, $\ket{\phi}=\sqrt{\frac{1}{2}+\frac{\sqrt{3}}{6}}\ket{0}+e^{i\frac{\pi}{4}}\sqrt{\frac{1}{2}-\frac{\sqrt{3}}{6}}\ket{1}$.

2. $\{+,+,-\}$, $\ket{\phi}=\sqrt{\frac{1}{2}-\frac{\sqrt{3}}{6}}\ket{0}+e^{i\frac{\pi}{4}}\sqrt{\frac{1}{2}+\frac{\sqrt{3}}{6}}\ket{1}$.

3. $\{+,-,+\}$, $\ket{\phi}=\sqrt{\frac{1}{2}+\frac{\sqrt{3}}{6}}\ket{0}+e^{-i\frac{\pi}{4}}\sqrt{\frac{1}{2}-\frac{\sqrt{3}}{6}}\ket{1}$.

4. $\{-,+,+\}$, $\ket{\phi}=\sqrt{\frac{1}{2}+\frac{\sqrt{3}}{6}}\ket{0}-e^{-i\frac{\pi}{4}}\sqrt{\frac{1}{2}-\frac{\sqrt{3}}{6}}\ket{1}$.

5. $\{+,-,-\}$, $\ket{\phi}=\sqrt{\frac{1}{2}-\frac{\sqrt{3}}{6}}\ket{0}+e^{-i\frac{\pi}{4}}\sqrt{\frac{1}{2}+\frac{\sqrt{3}}{6}}\ket{1}$.

6. $\{-,+,-\}$, $\ket{\phi}=\sqrt{\frac{1}{2}-\frac{\sqrt{3}}{6}}\ket{0}-e^{-i\frac{\pi}{4}}\sqrt{\frac{1}{2}+\frac{\sqrt{3}}{6}}\ket{1}$.

7. $\{-,-,+\}$, $\ket{\phi}=\sqrt{\frac{1}{2}+\frac{\sqrt{3}}{6}}\ket{0}-e^{i\frac{\pi}{4}}\sqrt{\frac{1}{2}-\frac{\sqrt{3}}{6}}\ket{1}$.

8. $\{-,-,-\}$, $\ket{\phi}=\sqrt{\frac{1}{2}-\frac{\sqrt{3}}{6}}\ket{0}-e^{i\frac{\pi}{4}}\sqrt{\frac{1}{2}+\frac{\sqrt{3}}{6}}\ket{1}$.

Thus according to the Schwarz inequality, when $\braket{H_j}=t\braket{\sigma}_j$ simultaneously with some constant $t$, the identities in the Equations are valid.

\section*{Acknowledgement}
This work is partially supported by the National Natural Science Foundation of China (12371135).


\begin{thebibliography}{28}%
\makeatletter
\providecommand \@ifxundefined [1]{%
 \@ifx{#1\undefined}
}%
\providecommand \@ifnum [1]{%
 \ifnum #1\expandafter \@firstoftwo
 \else \expandafter \@secondoftwo
 \fi
}%
\providecommand \@ifx [1]{%
 \ifx #1\expandafter \@firstoftwo
 \else \expandafter \@secondoftwo
 \fi
}%
\providecommand \natexlab [1]{#1}%
\providecommand \enquote  [1]{``#1''}%
\providecommand \bibnamefont  [1]{#1}%
\providecommand \bibfnamefont [1]{#1}%
\providecommand \citenamefont [1]{#1}%
\providecommand \href@noop [0]{\@secondoftwo}%
\providecommand \href [0]{\begingroup \@sanitize@url \@href}%
\providecommand \@href[1]{\@@startlink{#1}\@@href}%
\providecommand \@@href[1]{\endgroup#1\@@endlink}%
\providecommand \@sanitize@url [0]{\catcode `\\12\catcode `\$12\catcode
  `\&12\catcode `\#12\catcode `\^12\catcode `\_12\catcode `\%12\relax}%
\providecommand \@@startlink[1]{}%
\providecommand \@@endlink[0]{}%
\providecommand \url  [0]{\begingroup\@sanitize@url \@url }%
\providecommand \@url [1]{\endgroup\@href {#1}{\urlprefix }}%
\providecommand \urlprefix  [0]{URL }%
\providecommand \Eprint [0]{\href }%
\providecommand \doibase [0]{http://dx.doi.org/}%
\providecommand \selectlanguage [0]{\@gobble}%
\providecommand \bibinfo  [0]{\@secondoftwo}%
\providecommand \bibfield  [0]{\@secondoftwo}%
\providecommand \translation [1]{[#1]}%
\providecommand \BibitemOpen [0]{}%
\providecommand \bibitemStop [0]{}%
\providecommand \bibitemNoStop [0]{.\EOS\space}%
\providecommand \EOS [0]{\spacefactor3000\relax}%
\providecommand \BibitemShut  [1]{\csname bibitem#1\endcsname}%
\let\auto@bib@innerbib\@empty
%</preamble>
\bibitem [{\citenamefont {Heisenberg}(1927)}]{heisenberg1927anschaulichen}%
  \BibitemOpen
  \bibfield  {author} {\bibinfo {author} {\bibfnamefont {W.}~\bibnamefont
  {Heisenberg}},\ }\href@noop {} {\bibfield  {journal} {\bibinfo  {journal}
  {Zeitschrift f{\"u}r Physik}\ }\textbf {\bibinfo {volume} {43}},\ \bibinfo
  {pages} {172} (\bibinfo {year} {1927})}\BibitemShut {NoStop}%
\bibitem [{\citenamefont {Robertson}(1929)}]{robertson1929uncertainty}%
  \BibitemOpen
  \bibfield  {author} {\bibinfo {author} {\bibfnamefont {H.~P.}\ \bibnamefont
  {Robertson}},\ }\href@noop {} {\bibfield  {journal} {\bibinfo  {journal}
  {Physical Review}\ }\textbf {\bibinfo {volume} {34}},\ \bibinfo {pages} {163}
  (\bibinfo {year} {1929})}\BibitemShut {NoStop}%
\bibitem [{\citenamefont {Robertson}(1934)}]{robertson1934indeterminacy}%
  \BibitemOpen
  \bibfield  {author} {\bibinfo {author} {\bibfnamefont {H.}~\bibnamefont
  {Robertson}},\ }\href@noop {} {\bibfield  {journal} {\bibinfo  {journal}
  {Physical Review}\ }\textbf {\bibinfo {volume} {46}},\ \bibinfo {pages} {794}
  (\bibinfo {year} {1934})}\BibitemShut {NoStop}%
\bibitem [{\citenamefont {Maassen}\ and\ \citenamefont
  {Uffink}(1988)}]{maassen1988generalized}%
  \BibitemOpen
  \bibfield  {author} {\bibinfo {author} {\bibfnamefont {H.}~\bibnamefont
  {Maassen}}\ and\ \bibinfo {author} {\bibfnamefont {J.~B.}\ \bibnamefont
  {Uffink}},\ }\href@noop {} {\bibfield  {journal} {\bibinfo  {journal}
  {Physical Review Letters}\ }\textbf {\bibinfo {volume} {60}},\ \bibinfo
  {pages} {1103} (\bibinfo {year} {1988})}\BibitemShut {NoStop}%
\bibitem [{\citenamefont {Arthurs}\ and\ \citenamefont
  {Goodman}(1988)}]{arthurs1988quantum}%
  \BibitemOpen
  \bibfield  {author} {\bibinfo {author} {\bibfnamefont {E.}~\bibnamefont
  {Arthurs}}\ and\ \bibinfo {author} {\bibfnamefont {M.}~\bibnamefont
  {Goodman}},\ }\href@noop {} {\bibfield  {journal} {\bibinfo  {journal}
  {Physical Review Letters}\ }\textbf {\bibinfo {volume} {60}},\ \bibinfo
  {pages} {2447} (\bibinfo {year} {1988})}\BibitemShut {NoStop}%
\bibitem [{\citenamefont {Ozawa}(2004)}]{ozawa2004uncertainty}%
  \BibitemOpen
  \bibfield  {author} {\bibinfo {author} {\bibfnamefont {M.}~\bibnamefont
  {Ozawa}},\ }\href@noop {} {\bibfield  {journal} {\bibinfo  {journal} {Annals
  of Physics}\ }\textbf {\bibinfo {volume} {311}},\ \bibinfo {pages} {350}
  (\bibinfo {year} {2004})}\BibitemShut {NoStop}%
\bibitem [{\citenamefont {Busch}\ \emph {et~al.}(2007)\citenamefont {Busch},
  \citenamefont {Heinonen},\ and\ \citenamefont {Lahti}}]{busch2007heisenberg}%
  \BibitemOpen
  \bibfield  {author} {\bibinfo {author} {\bibfnamefont {P.}~\bibnamefont
  {Busch}}, \bibinfo {author} {\bibfnamefont {T.}~\bibnamefont {Heinonen}}, \
  and\ \bibinfo {author} {\bibfnamefont {P.}~\bibnamefont {Lahti}},\
  }\href@noop {} {\bibfield  {journal} {\bibinfo  {journal} {Physics Reports}\
  }\textbf {\bibinfo {volume} {452}},\ \bibinfo {pages} {155} (\bibinfo {year}
  {2007})}\BibitemShut {NoStop}%
\bibitem [{\citenamefont {Wehner}\ and\ \citenamefont
  {Winter}(2008)}]{wehner2008higher}%
  \BibitemOpen
  \bibfield  {author} {\bibinfo {author} {\bibfnamefont {S.}~\bibnamefont
  {Wehner}}\ and\ \bibinfo {author} {\bibfnamefont {A.}~\bibnamefont
  {Winter}},\ }\href@noop {} {\bibfield  {journal} {\bibinfo  {journal}
  {Journal of Mathematical Physics}\ }\textbf {\bibinfo {volume} {49}},\ \bibinfo 
  {pages} {062105} (\bibinfo {year} {2008})}\BibitemShut {NoStop}%
\bibitem [{\citenamefont {Wehner}\ and\ \citenamefont
  {Winter}(2010)}]{wehner2010entropic}%
  \BibitemOpen
  \bibfield  {author} {\bibinfo {author} {\bibfnamefont {S.}~\bibnamefont
  {Wehner}}\ and\ \bibinfo {author} {\bibfnamefont {A.}~\bibnamefont
  {Winter}},\ }\href@noop {} {\bibfield  {journal} {\bibinfo  {journal} {New
  Journal of Physics}\ }\textbf {\bibinfo {volume} {12}},\ \bibinfo {pages}
  {025009} (\bibinfo {year} {2010})}\BibitemShut {NoStop}%
\bibitem [{\citenamefont {Watanabe}\ \emph {et~al.}(2011)\citenamefont
  {Watanabe}, \citenamefont {Sagawa},\ and\ \citenamefont
  {Ueda}}]{watanabe2011uncertainty}%
  \BibitemOpen
  \bibfield  {author} {\bibinfo {author} {\bibfnamefont {Y.}~\bibnamefont
  {Watanabe}}, \bibinfo {author} {\bibfnamefont {T.}~\bibnamefont {Sagawa}}, \
  and\ \bibinfo {author} {\bibfnamefont {M.}~\bibnamefont {Ueda}},\ }\href@noop
  {} {\bibfield  {journal} {\bibinfo  {journal} {Physical Review A}\ }\textbf {\bibinfo {volume} {84}},\ \bibinfo
  {pages} {042121} (\bibinfo {year} {2011})}\BibitemShut {NoStop}%
\bibitem [{\citenamefont {Huang}(2012)}]{huang2012variance}%
  \BibitemOpen
  \bibfield  {author} {\bibinfo {author} {\bibfnamefont {Y.}~\bibnamefont
  {Huang}},\ }\href@noop {} {\bibfield  {journal} {\bibinfo  {journal}
  {Physical Review A}\ }\textbf
  {\bibinfo {volume} {86}},\ \bibinfo {pages} {024101} (\bibinfo {year}
  {2012})}\BibitemShut {NoStop}%
\bibitem [{\citenamefont {Chen}\ \emph {et~al.}(2016)\citenamefont {Chen},
  \citenamefont {Cao}, \citenamefont {Fei},\ and\ \citenamefont
  {Long}}]{chen2016variance}%
  \BibitemOpen
  \bibfield  {author} {\bibinfo {author} {\bibfnamefont {B.}~\bibnamefont
  {Chen}}, \bibinfo {author} {\bibfnamefont {N.-P.}\ \bibnamefont {Cao}},
  \bibinfo {author} {\bibfnamefont {S.-M.}\ \bibnamefont {Fei}}, \ and\
  \bibinfo {author} {\bibfnamefont {G.-L.}\ \bibnamefont {Long}},\ }\href@noop
  {} {\bibfield  {journal} {\bibinfo  {journal} {Quantum Information
  Processing}\ }\textbf {\bibinfo {volume} {15}},\ \bibinfo {pages} {3909}
  (\bibinfo {year} {2016})}\BibitemShut {NoStop}%
\bibitem [{\citenamefont {Li}\ and\ \citenamefont
  {Qiao}(2016)}]{li2016equivalence}%
  \BibitemOpen
  \bibfield  {author} {\bibinfo {author} {\bibfnamefont {J.-L.}\ \bibnamefont
  {Li}}\ and\ \bibinfo {author} {\bibfnamefont {C.-F.}\ \bibnamefont {Qiao}},\
  }\href@noop {} {\bibfield  {journal} {\bibinfo  {journal} {Journal of Physics
  A: Mathematical and Theoretical}\ }\textbf {\bibinfo {volume} {50}},\
  \bibinfo {pages} {03LT01} (\bibinfo {year} {2016})}\BibitemShut {NoStop}%
\bibitem [{\citenamefont {Coles}\ \emph {et~al.}(2017)\citenamefont {Coles},
  \citenamefont {Berta}, \citenamefont {Tomamichel},\ and\ \citenamefont
  {Wehner}}]{coles2017entropic}%
  \BibitemOpen
  \bibfield  {author} {\bibinfo {author} {\bibfnamefont {P.~J.}\ \bibnamefont
  {Coles}}, \bibinfo {author} {\bibfnamefont {M.}~\bibnamefont {Berta}},
  \bibinfo {author} {\bibfnamefont {M.}~\bibnamefont {Tomamichel}}, \ and\
  \bibinfo {author} {\bibfnamefont {S.}~\bibnamefont {Wehner}},\ }\href@noop {}
  {\bibfield  {journal} {\bibinfo  {journal} {Reviews of Modern Physics}\
  }\textbf {\bibinfo {volume} {89}},\ \bibinfo {pages} {015002} (\bibinfo
  {year} {2017})}\BibitemShut {NoStop}%
\bibitem [{\citenamefont {Miyadera}\ and\ \citenamefont
  {Imai}(2008)}]{miyadera2008heisenberg}%
  \BibitemOpen
  \bibfield  {author} {\bibinfo {author} {\bibfnamefont {T.}~\bibnamefont
  {Miyadera}}\ and\ \bibinfo {author} {\bibfnamefont {H.}~\bibnamefont
  {Imai}},\ }\href@noop {} {\bibfield  {journal} {\bibinfo  {journal} {Physical
  Review A}\ }\textbf {\bibinfo
  {volume} {78}},\ \bibinfo {pages} {052119} (\bibinfo {year}
  {2008})}\BibitemShut {NoStop}%
\bibitem [{\citenamefont {Dodonov}(2019)}]{dodonov2019uncertainty}%
  \BibitemOpen
  \bibfield  {author} {\bibinfo {author} {\bibfnamefont {V.}~\bibnamefont
  {Dodonov}},\ }in\ \href@noop {} {\emph {\bibinfo {booktitle} {Journal of
  Physics: Conference Series}}},\ \textbf{\bibinfo {volume} {1194}}, \bibinfo{pages} {012028} 
  (\bibinfo {year} {2019})  \BibitemShut {NoStop}%
\bibitem [{\citenamefont {Liang}\ \emph {et~al.}(2024)\citenamefont {Liang},
  \citenamefont {Li},\ and\ \citenamefont {Fei}}]{liang2024signifying}%
  \BibitemOpen
  \bibfield  {author} {\bibinfo {author} {\bibfnamefont {X.-B.}\ \bibnamefont
  {Liang}}, \bibinfo {author} {\bibfnamefont {B.}~\bibnamefont {Li}}, \ and\
  \bibinfo {author} {\bibfnamefont {S.-M.}\ \bibnamefont {Fei}},\ }\href@noop
  {} {\bibfield  {journal} {\bibinfo  {journal} {Science China Physics,
  Mechanics \& Astronomy}\ }\textbf {\bibinfo {volume} {67}},\ \bibinfo {pages}
  {290311} (\bibinfo {year} {2024})}\BibitemShut {NoStop}%
\bibitem [{\citenamefont {Einstein}\ \emph {et~al.}(1935)\citenamefont
  {Einstein}, \citenamefont {Podolsky},\ and\ \citenamefont
  {Rosen}}]{einstein1935can}%
  \BibitemOpen
  \bibfield  {author} {\bibinfo {author} {\bibfnamefont {A.}~\bibnamefont
  {Einstein}}, \bibinfo {author} {\bibfnamefont {B.}~\bibnamefont {Podolsky}},
  \ and\ \bibinfo {author} {\bibfnamefont {N.}~\bibnamefont {Rosen}},\
  }\href@noop {} {\bibfield  {journal} {\bibinfo  {journal} {Physical Review}\
  }\textbf {\bibinfo {volume} {47}},\ \bibinfo {pages} {777} (\bibinfo {year}
  {1935})}\BibitemShut {NoStop}%
\bibitem [{\citenamefont {Horodecki}\ \emph {et~al.}(2009)\citenamefont
  {Horodecki}, \citenamefont {Horodecki}, \citenamefont {Horodecki},\ and\
  \citenamefont {Horodecki}}]{horodecki2009quantum}%
  \BibitemOpen
  \bibfield  {author} {\bibinfo {author} {\bibfnamefont {R.}~\bibnamefont
  {Horodecki}}, \bibinfo {author} {\bibfnamefont {P.}~\bibnamefont
  {Horodecki}}, \bibinfo {author} {\bibfnamefont {M.}~\bibnamefont
  {Horodecki}}, \ and\ \bibinfo {author} {\bibfnamefont {K.}~\bibnamefont
  {Horodecki}},\ }\href@noop {} {\bibfield  {journal} {\bibinfo  {journal}
  {Reviews of Modern Physics}\ }\textbf {\bibinfo {volume} {81}},\ \bibinfo
  {pages} {865} (\bibinfo {year} {2009})}\BibitemShut {NoStop}%
\bibitem [{\citenamefont {Bell}(1964)}]{bell1964einstein}%
  \BibitemOpen
  \bibfield  {author} {\bibinfo {author} {\bibfnamefont {J.~S.}\ \bibnamefont
  {Bell}},\ }\href@noop {} {\bibfield  {journal} {\bibinfo  {journal} {Physics
  Physique Fizika}\ }\textbf {\bibinfo {volume} {1}},\ \bibinfo {pages} {195}
  (\bibinfo {year} {1964})}\BibitemShut {NoStop}%
\bibitem [{\citenamefont {Cirel'son}(1980)}]{cirel1980quantum}%
  \BibitemOpen
  \bibfield  {author} {\bibinfo {author} {\bibfnamefont {B.~S.}\ \bibnamefont
  {Cirel'son}},\ }\href@noop {} {\bibfield  {journal} {\bibinfo  {journal}
  {Letters in Mathematical Physics}\ }\textbf {\bibinfo {volume} {4}},\
  \bibinfo {pages} {93} (\bibinfo {year} {1980})}\BibitemShut {NoStop}%
\bibitem [{\citenamefont
  {Landau}(1987{\natexlab{a}})}]{landau1987experimental}%
  \BibitemOpen
  \bibfield  {author} {\bibinfo {author} {\bibfnamefont {L.~J.}\ \bibnamefont
  {Landau}},\ }\href@noop {} {\bibfield  {journal} {\bibinfo  {journal}
  {Letters in Mathematical Physics}\ }\textbf {\bibinfo {volume} {14}},\
  \bibinfo {pages} {33} (\bibinfo {year} {1987}{\natexlab{a}})}\BibitemShut
  {NoStop}%
\bibitem [{\citenamefont {Landau}(1987{\natexlab{b}})}]{landau1987violation}%
  \BibitemOpen
  \bibfield  {author} {\bibinfo {author} {\bibfnamefont {L.~J.}\ \bibnamefont
  {Landau}},\ }\href@noop {} {\bibfield  {journal} {\bibinfo  {journal}
  {Physics Letters A}\ }\textbf {\bibinfo {volume} {120}},\ \bibinfo {pages}
  {54} (\bibinfo {year} {1987}{\natexlab{b}})}\BibitemShut {NoStop}%
\bibitem [{\citenamefont {Hofmann}\ and\ \citenamefont
  {Takeuchi}(2003)}]{hofmann2003violation}%
  \BibitemOpen
  \bibfield  {author} {\bibinfo {author} {\bibfnamefont {H.~F.}\ \bibnamefont
  {Hofmann}}\ and\ \bibinfo {author} {\bibfnamefont {S.}~\bibnamefont
  {Takeuchi}},\ }\href@noop {} {\bibfield  {journal} {\bibinfo  {journal}
  {Physical Review A}\ }\textbf {\bibinfo {volume} {68}},\ \bibinfo {pages}
  {032103} (\bibinfo {year} {2003})}\BibitemShut {NoStop}%
\bibitem [{\citenamefont {Wolf}\ \emph {et~al.}(2009)\citenamefont {Wolf},
  \citenamefont {Perez-Garcia},\ and\ \citenamefont
  {Fernandez}}]{wolf2009measurements}%
  \BibitemOpen
  \bibfield  {author} {\bibinfo {author} {\bibfnamefont {M.}~\bibnamefont
  {Wolf}}, \bibinfo {author} {\bibfnamefont {D.}~\bibnamefont {Perez-Garcia}},
  \ and\ \bibinfo {author} {\bibfnamefont {C.}~\bibnamefont {Fernandez}},\
  }\href@noop {} {\bibfield  {journal} {\bibinfo  {journal} {Physical Review
  Letters}\ }\textbf {\bibinfo {volume} {103}},\ \bibinfo {pages} {230402}
  (\bibinfo {year} {2009})}\BibitemShut {NoStop}%
\bibitem [{\citenamefont {Oppenheim}\ and\ \citenamefont
  {Wehner}(2010)}]{oppenheim2010uncertainty}%
  \BibitemOpen
  \bibfield  {author} {\bibinfo {author} {\bibfnamefont {J.}~\bibnamefont
  {Oppenheim}}\ and\ \bibinfo {author} {\bibfnamefont {S.}~\bibnamefont
  {Wehner}},\ }\href@noop {} {\bibfield  {journal} {\bibinfo  {journal}
  {Science}\ }\textbf {\bibinfo {volume} {330}},\ \bibinfo {pages} {1072}
  (\bibinfo {year} {2010})}\BibitemShut {NoStop}%
\bibitem [{\citenamefont {Ma}\ \emph {et~al.}(2017)\citenamefont {Ma},
  \citenamefont {Chen}, \citenamefont {Liu}, \citenamefont {Wang},
  \citenamefont {Ye}, \citenamefont {Kong}, \citenamefont {Shi}, \citenamefont
  {Fei},\ and\ \citenamefont {Du}}]{ma2017experimental}%
  \BibitemOpen
  \bibfield  {author} {\bibinfo {author} {\bibfnamefont {W.}~\bibnamefont
  {Ma}}, \bibinfo {author} {\bibfnamefont {B.}~\bibnamefont {Chen}}, \bibinfo
  {author} {\bibfnamefont {Y.}~\bibnamefont {Liu}}, \bibinfo {author}
  {\bibfnamefont {M.}~\bibnamefont {Wang}}, \bibinfo {author} {\bibfnamefont
  {X.}~\bibnamefont {Ye}}, \bibinfo {author} {\bibfnamefont {F.}~\bibnamefont
  {Kong}}, \bibinfo {author} {\bibfnamefont {F.}~\bibnamefont {Shi}}, \bibinfo
  {author} {\bibfnamefont {S.-M.}\ \bibnamefont {Fei}}, \ and\ \bibinfo
  {author} {\bibfnamefont {J.}~\bibnamefont {Du}},\ }\href@noop {} {\bibfield
  {journal} {\bibinfo  {journal} {Physical Review Letters}\ }\textbf {\bibinfo
  {volume} {118}},\ \bibinfo {pages} {180402} (\bibinfo {year}
  {2017})}\BibitemShut {NoStop}%
\bibitem [{\citenamefont {Watrous}(2018)}]{watrous2018theory}%
  \BibitemOpen
  \bibfield  {author} {\bibinfo {author} {\bibfnamefont {J.}~\bibnamefont
  {Watrous}},\ }\href@noop {} {\emph {\bibinfo {title} {The theory of quantum
  information}}}\ (\bibinfo  {publisher} {Cambridge university press},\
  \bibinfo {year} {2018})\BibitemShut {NoStop}%
\end{thebibliography}
\end{document}